\documentstyle[twocolumn,floats,epsf,epsfig,ifthen,aps,prl]{revtex}

\begin{document}

\title{Quantum Nondemolition Squeezing of a Nanomechanical Resonator}

\author{Rusko Ruskov\footnote{On leave from
INRNE, Sofia BG-1784, Bulgaria}$^{\rm 1}$,
Keith Schwab$^{\rm 2}$, and Alexander N. Korotkov$^{\rm 1}$}

\address{$^{\rm 1}$Department of Electrical Engineering, University of
California, Riverside, CA 92521}

\address{$^{\rm 2}$Laboratory for Physical Sciences, 8050 Greenmead Drive, College Park,
MD 20740}

\date{\today}

\maketitle

\begin{abstract}
We show that the nanoresonator position can be squeezed
significantly below the ground state level by measuring the nanoresonator
with a quantum point contact or a single-electron transistor and
applying a periodic voltage across the detector. The mechanism of
squeezing is basically a generalization of quantum nondemolition
measurement of an oscillator to the case of continuous measurement
by a weakly coupled detector. The quantum feedback is necessary
to prevent the ``heating'' due to measurement back-action.
We also discuss a procedure of experimental verification of the squeezed
state.
\end{abstract}

\section{Introduction}

Quantum nondemolition (QND) measurements
\cite{BraginskyKhalili,Braginsky2,Thorne,CavesRevModPhys} were proposed
as a way to overcome the so-called standard quantum limit for
the measurement sensitivity, which arises due to quantum back-action
of a detector. The general idea of a QND measurement is to avoid
measuring (or obtaining any information on) the magnitude conjugated
to the magnitude of interest, and therefore to avoid the corresponding
back-action. An important implementation of this idea is the ``stroboscopic''
measurement of an oscillator position \cite{Braginsky2,Thorne}.
Suppose the position $x_1$ is measured (instantaneously) with a finite
precision $\Delta x$, which necessarily disturbs the momentum according
to the Heisenberg
uncertainty principle $\Delta p \geq \hbar /2\Delta x$. Normally this
momentum change would affect the result of the next position measurement $x_2$
and would limit the accuracy for the position difference $x_2-x_1$,
leading to the standard quantum limit for this magnitude.
However, if the second measurement
is performed exactly one oscillation period after the first one, the
oscillator returns to its initial state, and therefore the momentum change
does not affect the accuracy of $x_2-x_1$ measurement. Such stroboscopic
measurement gives no information related to the momentum, and this is exactly
the reason why the effect of quantum back-action is avoided
\cite{BraginskyKhalili,Braginsky2,Thorne,CavesRevModPhys}.

        In terms of applications, the QND measurements have been mainly
discussed in relation to measurement of very weak classical forces,
in particular gravitational waves (see, e.g., \cite{Kimble,Braginsky-2003}).
Recently the idea of QND measurements has been also applied to solid-state
mesoscopic structures (see, e.g., \cite{Averin,Bulaevskii}). Among other
recent developments (total number of papers on QND measurements is over
half a thousand) let us mention the experiment on atomic spin-squeezing
using the QND
measurement and real-time quantum feedback \cite{Geremia}. In this paper
we discuss squeezing of a nanomechanical resonator using the QND measurement
and quantum feedback.

        Recent advances in fabrication of nanomechanical resonators
\cite{ClelandRoukes,Craighead,Roukes-1GHz,Cleland-2003,Schwab-QL}
make possible the direct observation of their quantum behavior in the nearest
future.
For the resonator frequency $\omega_0/2\pi$ exceeding 1 GHz
\cite{Roukes-1GHz} the condition $T < \hbar\omega_0$ (we use $k_B=1$)
is satisfied at
temperature $T$ below $\sim 50$ mK. Even in the case $T \gg \hbar\omega_0$
the quantum behavior is in principle observable  \cite{BraginskyKhalili}
when $T\tau/Q < \hbar/2 $, where  $Q$ is the resonator
quality factor and $\tau$ is the typical observation time.
This condition can be well satisfied even for MHz-range resonators
for $\tau$ comparable to the oscillation period, i.e.\ if
we can monitor the oscillations with the measurement bandwidth on the
order of $\omega_0$, as in Ref.\ \cite{Schwab-QL}.
There is a rapid
experimental progress in monitoring the oscillating position of a
nanoresonator using radio-frequency single-electron transistor (RF-SET)
\cite{Cleland-2003,Schwab-QL} or quantum point contact (QPC)
\cite{Cleland-QPC} (at present RF-SET seems to be much more
efficient).
In particular, the position measurement accuracy $\Delta x$ within
the factor 5.8 from the standard quantum limit $\Delta x_0$ has been
demonstrated \cite{Schwab-QL}; here $\Delta x_0 = \sqrt{\hbar /2m\omega_0}$
is the width (standard deviation) of the ground state of the oscillator
with mass $m$. Measurement of the nanoresonator position by RF-SET or QPC
has also received a significant theoretical attention
\cite{Blencowe,MozyrskyMartin,ArmourBlencoweSchwab,Smirnov,Schwab,Knobel,%
Hopkins}.
The process of measurement transfers the energy from the
detector to the nanoresonator leading to its ``heating''
\cite{Blencowe,MozyrskyMartin}. A possible way to prevent such heating
is using the quantum feedback control of the nanoresonator \cite{Hopkins}
(another idea for cooling has been proposed in Ref.\ \cite{ShnirmanMartin}).

The quantum feedback of mesoscopic solid-state systems is a relatively
new subject \cite{Ruskov-fb}, though in quantum optics the quantum feedback
has been proposed more than a decade ago \cite{Wiseman} and has been
already realized experimentally \cite{Geremia}. The feedback analyzed
in Ref.\ \cite{Hopkins} assumes continuous monitoring of the nanoresonator
position with constant ``strength'' of measurement
and  allows cooling of the nanoresonator practically down to
the ground state. However, it does not allow squeezing of the nanoresonator
state (below $\Delta x_0$), that would be important, for example, for the
ultrasensitive measurement of a force acting on the nanoresonator.
(More correctly, squeezing in Ref.\ \cite{Hopkins} is possible only in the
unrealistic case of a strong coupling between the nanoresonator and detector.)
In this paper we analyze the case of the nanoresonator monitoring
with the measurement strength modulated in time (for example, modulating
the bias voltage of the QPC or RF-SET), basically applying the
idea of stroboscopic QND measurements \cite{Braginsky2,Thorne} to the
nanoresonator. We show that significant squeezing of the nanoresonator state
can be achieved when the modulation frequency $\omega$ is close to
$2\omega_0/n$, $n=1, 2, \ldots$, similar to the usual QND result.
The difference from the case of Refs.\ \cite{Braginsky2,Thorne} though is
that we consider continuous in time measurement and assume weak coupling
between the nanoresonator and the detector; also, we use the quantum
feedback to prevent the heating due to measurement. We would also like
to notice the difference between our proposal and squeezing of a
nanoresonator proposed in Ref.\ \cite{Blencowe-sq}, which is a scaled-down
version of the classical thermomechanical noise squeezing \cite{RugarGrutter}
using the modulation of the resonator spring constant.

\section{System and model}

    For simplicity we consider the nanoresonator measured by
the low-transparency QPC (though our results are applicable to the RF-SET
as well)
and the system Hamiltonian is
        \begin{equation}
  {\cal H} = {\cal H}_{0} + {\cal H}_{det} + {\cal H}_{int} + {\cal H}_{fb},
        \label{Hamiltonian}
        \end{equation}
where the first term describes the oscillator (nanoresonator):
        \begin{equation}
  {\cal H}_{0} = \frac{\hat{p}^2}{2 m} + \frac{m \omega_0^2 \hat{x}^2}{2}
        \label{oscillator}
        \end{equation}
($\hat{p}$ and $\hat{x}$ being the momentum and position operators),
the last term
        \begin{equation}
  {\cal H}_{fb} = - {\cal F} \hat{x}
        \end{equation}
describes the feedback control of the nanoresonator by applying the force
${\cal F} (t)$, while ${\cal H}_{det}$ and ${\cal H}_{int}$ correspond to
the detector and its interaction with the nanoresonator
similar to Refs.\ \cite{Gurvitz,MozyrskyMartin}:
        \begin{eqnarray}
   &&
{\cal H}_{det} = \sum\nolimits_l E_l a_l^\dagger a_l +
\sum\nolimits_r E_r a_r^\dagger a_r
      \nonumber \\
 &&\hspace{1.3cm}
+\sum\nolimits_{l,r} ( M \, a_l^\dagger a_r+ \mbox{H.c.}) \, ,
\label{Hdet}\\
&& {\cal H}_{int}= \sum\nolimits_{l,r} ( \Delta M \, \hat{x} \,
a_l^\dagger a_r + \mbox{H.c.} ) \, .
        \label{Hint}
        \end{eqnarray}
Here $a^\dagger_{l,r}$ and $a_{l,r}$ are the creation and annihilation
operators for two electrodes of the QPC, and for simplicity we assume
no relative phase between the tunneling amplitudes $M$ and $\Delta M$
(taking this phase into account is simple \cite{Kor-Av,GoanMilburn,Kor-nonid},
but makes the formalism significantly lengthier).
For a given position state $|x\rangle$ of the oscillator, the average
detector current is $I_x=2\pi |M+\Delta M x|^2 \rho_l\rho_r e^2V/\hbar$,
where $V$ is the QPC voltage which may vary in time with frequency $\omega$
comparable to $\omega_0$, $e$ is the electron
charge, and $\rho_{l,r}$ are the densities of states in the electrodes.

        We assume a weak response of the detector, $|I_x-I_{x'}| \ll
I_x+I_{x'}$, and therefore the linear dependence of the detector
current on the measured position
        \begin{equation}
I_x = I_0 + k x,
        \label{linearity}
        \end{equation}
also we neglect the dependence on $x$ of the detector current spectral
density $S_I$ which is assumed to be flat in the frequency range of
interest. Because the voltage $V$ varies in time, $I_0$, $k$, $I_x$,
and $S_I$ also depend on time, which will be taken into account explicitly
in the next Section.

        Somewhat similar to the case of qubit measurement
\cite{Kor-99-01},
we define the dimensionless (time-dependent) coupling as
        \begin{equation}
{\cal C}=\frac{\hbar k^2}{S_I m \omega_0^2} ,
        \label{coupling}
        \end{equation}
which can also be expressed as ${\cal C}=4/\omega_0\tau_m$, where
$\tau_m = 2S_I/(k \Delta x_0)^2$ is the ``measurement'' time which would
be necessary
to distinguish (with signal-to-noise ratio of 1) two position states
separated by the ground state width $\Delta x_0 = \sqrt{\hbar /2m\omega_0}$.
We will mainly consider the case of weak coupling, ${\cal C}\ll 1$,
which corresponds to a realistic experimental situation. As an example,
${\cal C}$ is on the order of $10^{-3}$ for the parameters of experiment
\cite{Schwab-QL}.

        To describe the dynamics of the quantum measurement process,
we apply the Bayesian approach \cite{Kor-99-01}, which is practically
equivalent to the approach of quantum trajectories used, e.g., in Refs.\
\cite{Hopkins,Wiseman,GoanMilburn,DohertyJacobs}. Therefore we need
to use the usual assumptions of the Bayesian approach; in particular,
we assume that the internal dynamics of the detector is much faster
than the oscillator dynamics (this requires $eV \gg \hbar\omega_0$),
and we assume quasicontinuous detector current (which requires
$I_0/e \gg \omega_0$
and even stronger inequality $k\Delta x_0/e \gg \omega_0$).

        Applying the Bayesian approach to our system, we derive
(derivation will be presented elsewhere) the following equation
for the evolution of oscillator density matrix $\rho$ in $x$-basis
(in Stratonovich form):
        \begin{eqnarray}
&& \hspace{0.3cm}
\dot{\rho}(x,x') = \frac{-i}{\hbar} [{\cal H}_0+{\cal H}_{fb},
        \rho ]_{x,x'} +
\rho(x,x')\frac{1}{S_I}
\nonumber\\
&& \hspace{0.5cm}
\times \left\{ I(t) \left(I_x + I_{x'} - 2\langle I \rangle
\right) -
\left( \frac{I_x^2 + I_{x'}^2}{2} - \langle I^2 \rangle \right)
\right\} ,
        \label{meas-Start}
        \end{eqnarray}
where the first term is a usual evolution due to ${\cal H}_0+{\cal H}_{fb}$,
while the second term describes the evolution due to measurement, and
therefore
depends on the noisy detector current $I(t)$; we introduced notations
$\langle I\rangle = \int I_x \rho (x,x) dx$ and
$\langle I^2\rangle = \int I_x^2 \rho (x,x) dx$.
Notice that Eq.\ (\ref{meas-Start}) actually does not require the current
linearity (\ref{linearity}). Also notice that Eq. (\ref{meas-Start})
coincides with the similar equation for the case of arbitrary number
of entangled qubits measured by an ideal detector \cite{Kor-nonid},
if $x$ is replaced by the index corresponding to the state of qubits.

        Equation (\ref{meas-Start}) allows us to monitor the oscillator
density matrix $\rho$ using the measurement record $I(t)$, while for
simulations $I(t)$ may be replaced with
        \begin{equation}
I(t) = \langle I\rangle + \xi(t),
        \label{noisy-curr}
\end{equation}
where $\xi(t)$ is a white noise with spectral density $S_I$.

        Translating Eq.\ (\ref{meas-Start}) from Stratonovich into It\^o
form, using current linearity (\ref{linearity}), and taking into account
quantum efficiency $\eta$ of the detector \cite{Kor-99-01}
(QPC is an ideal detector, $\eta =1$, while $\eta <1$ can be used for
the RF-SET as a detector), we obtain
\begin{eqnarray}
&& \hspace{0.08cm}
\dot{\rho}(x,x') = \frac{- i}{\hbar} [{\cal H}_0 +{\cal H}_{fb},\rho ]_{x,x'}
 -\frac{k^2}{4S_I \eta} (x-x')^2 \, \rho(x,x')
\nonumber\\
&& \hspace{1.6cm}
+ \frac{k}{S_I} (x+x'-2\langle x\rangle ) \, \rho(x,x') \, \xi(t) ,
\label{meas-Ito-linear}
\end{eqnarray}
where $\langle x\rangle = \int x\, \rho (x,x)dx$. Equation
(\ref{meas-Ito-linear}) is similar to equations derived in many publications
(e.g., in Refs.\ \cite{Mensky,GWM,DohertyJacobs,Hopkins})
for measurement of a mechanical oscillator.
    Averaging Eq.\ (\ref{meas-Ito-linear}) over the measurement record
$I(t)$ eliminates the last term of Eq.\ (\ref{meas-Ito-linear}) [in It\^o
form averaging over the noise $\xi (t)$ is equivalent to using $\xi (t) =0$],
and leads to the ensemble averaged equation derived in even
larger number of papers, including, e.g., Ref.\ \cite{MozyrskyMartin}.
Notice that the
second (decoherence) term in Eq.\ (\ref{meas-Ito-linear}) can also be
rewritten in a standard double-commutator form (see, e.g., Refs.\
\cite{Lindblad,Leggett,CavesMilburn,GWM,ZurekHabibPaz-prl,MozyrskyMartin,%
DohertyJacobs,Hopkins})
since $(x-x')^2\rho (x,x') = [\hat{x}, [\hat{x},\rho]]_{x,x'}$.
The nanoresonator evolution described by Eq.\ (\ref{meas-Ito-linear})
does not depend on the environment temperature because we essentially
assume large (infinite) quality factor of the nanoresonator, so that
the interaction with the thermal bath is much weaker than interaction
with the detector which has infinite by assumption effective temperature
$T_d \sim eV/2$ \cite{MozyrskyMartin,Blencowe}.

\section{QND squeezing of the nanoresonator}

\subsection{Modulation of the measurement strength}

        We assume periodic modulation of the voltage across the QPC
detector, $V = f(t) V_0$, which leads to the corresponding modulation
of the parameters in Eqs.\ (\ref{meas-Start}) and  (\ref{meas-Ito-linear}):
    \begin{equation}
k=f(t)k_0, \hspace{0.2cm}
I_x= f(t) (I_{00}+k_0 x), \hspace{0.2cm}
S_I =|f(t)| S_0 ,
    \end{equation}
while quantum efficiency $\eta$ is assumed constant (in general case
$f(t)$ may become negative). Notice that
the noise $\xi (t)$ has implicit time dependence because of modulated
in time spectral density $S_I$. The dimensionless coupling is modulated
as ${\cal C}=|f(t)|{\cal C}_0$.

        In this paper we will consider two types of modulation with frequency
$\omega$: harmonic modulation with 100\% depth
        \begin{equation}
f(t)= (1+\cos \omega t)/2
        \label{harm-mod}
        \end{equation}
and the square-wave (stroboscopic) modulation with pulse width $\delta t$
        \begin{equation}
f(t) = \left\{ \begin{array}{l}  1, \,\,\,
|t-j\times 2\pi /\omega |\leq \delta t/2, \,\, j=1, 2, \ldots
\\ 0, \,\,\, \mbox{otherwise} \, .
        \end{array} \right.
        \label{strob-mod}
        \end{equation}
Notice that $|f(t)|\leq 1$, so $k_0$ and ${\cal C}_0$ correspond to the
maximum coupling.
        Since $f(t)$ reaches zero in both types of modulation, the conditions
$eV\gg \hbar \omega_0$ and $k\Delta x_0/e\gg \omega_0$ required for the
Bayesian formalism are violated during a fraction of the modulation period.
However, the Bayesian formalism is trivially correct at $V=0$ when the
nanoresonator is not measured (neglecting small remaining coupling
\cite{MozyrskyMartin}). So, assuming that the intermediate regime for
which the Bayesian formalism is inapplicable, is realized during only a short
fraction of the modulation period, we still use Eq.\ (\ref{meas-Ito-linear})
for the analysis.

\subsection{Simplified equations for the Gaussian states}

Following Refs.\ \cite{HalliwellZoupas,BreslinMilburn,DohertyJacobs,%
Hopkins}, we assume that the oscillator state is Gaussian \cite{Gardiner}:
\begin{eqnarray}
&& \hspace{0.3cm}
\rho(x,x')=\frac{1}{\sqrt{2\pi D_x}} \,
\exp{\left[-\frac{(\frac{x+x'}{2}-\langle x\rangle )^2}{2 D_x}\right]}
\nonumber\\
&& \hspace{0.5cm}
\times \exp{\left[-\frac{(x-x')^2}{8 D_x} \,
\frac{(D_x D_p - D_{xp}^2)}{\hbar^2/4} \right]}
\nonumber\\
&& \hspace{0.5cm}
\times \exp{\left[ i (x-x') \left( \frac{\langle p\rangle }{\hbar} +
(\frac{x+x'}{2}-\langle x\rangle )\,
\frac{D_{xp}}{\hbar D_x} \right) \right]}
        \label{Gaussian-den-mat}
        \end{eqnarray}
and therefore is characterized by only five parameters:
average position $\langle x\rangle = \langle \hat{x}\rangle$ and
momentum $\langle p\rangle =\langle \hat{p}\rangle$,
their variances $D_x=\langle \hat{x}^2\rangle -\langle \hat{x}\rangle^2$
and $D_p=\langle \hat{p}^2\rangle -\langle \hat{p}\rangle^2$,
and the correlation $D_{xp} =\langle \hat{x}\hat{p}+\hat{p}\hat{x}\rangle /2
-\langle\hat{x}\rangle \langle\hat{p}\rangle$. These parameters satisfy
the generalized Heisenberg inequality \cite{Man'ko}
        \begin{equation}
D_x D_p - D_{xp}^2 \geq \hbar^2/4 ,
        \label{uncertain}
        \end{equation}
which reaches the lower bound for the pure states.
The assumption of the Gaussian state can be justified by the fact that
a Gaussian state remains Gaussian in the process of continuous
measurement \cite{BreslinMilburn} (we have checked this statement for
nonideal detectors including ``asymmetric'' detectors and for varying
in time strength of measurement) and by the fact that the thermal state
(natural initial condition) is Gaussian \cite{Gardiner}.

        For Gaussian states Eq.\ (\ref{meas-Ito-linear}) transforms into
equations
        \begin{eqnarray}
&& \hspace{0.2cm}
\dot{\langle x\rangle } =\frac{\langle p\rangle}{m}
+ \frac{2k_0}{S_0}\ \mbox{sgn}[f(t)]\, D_x\, \xi(t) \, ,
\label{eq-m-x}\\
&& \hspace{0.2cm}
\dot{\langle p\rangle }=-m \omega_0^2 \langle x\rangle
+ \frac{2k_0}{S_0}\ \mbox{sgn}[f(t)]\, D_{xp}\, \xi(t) +{\cal F} \, ,
\label{eq-m-p}\\
&& \hspace{0.2cm}
\dot{D}_{x}=\frac{2}{m} D_{xp}
- \frac{2k_0^2}{S_0}\ |f(t)|\, D_x^2 \, ,
\label{eq-mDx}\\
&& \hspace{0.2cm}
\dot{D}_{p}=-2m \omega_0^2 D_{xp} +
\frac{k_0^2 \hbar^2}{2S_0 \eta}\ |f(t)|- \frac{2k_0^2}{S_0}\ |f(t)|\,
        D_{xp}^2  \, ,
\label{eq-mDp}\\
&& \hspace{0.2cm}
\dot{D}_{xp}=\frac{1}{m} D_{p} -m \omega_0^2 D_{x}-
\frac{2k_0^2}{S_0}\ |f(t)|\, D_x D_{xp} \, ,
\label{eq-mDxp}
\end{eqnarray}
        which practically coincide with the equations derived in Refs.\
\cite{HalliwellZoupas,DohertyJacobs,Hopkins}, except for the time dependent
$f(t)$. It is interesting to notice that while Eq.\ (\ref{meas-Ito-linear})
is a nonlinear stochastic equation, for which the Stratonovich and It\^o
forms are significantly different, there is no such difference for
Eqs.\ (\ref{eq-m-x})--(\ref{eq-mDxp}), so they can be treated as simple
ordinary differential equations.

        Notice that the equations for $D_x$, $D_p$, and $D_{xp}$ do not
depend on noise $\xi (t)$ and feedback force ${\cal F}$, and are
decoupled from the remaining equations.
Therefore the evolution of the ``wave packet width'' $\sqrt{D_x}$ is
deterministic. Analyzing the possibility to squeeze the nanoresonator state,
we will consider separately the squeezing of the packet width
and the contribution $D_{\langle x\rangle}$
to the total position variance due to fluctuating position of the packet
center $\langle x\rangle$. As will be discussed below, $D_x$ may be made
significantly
smaller than the ground state variance $\Delta x_0^2$ using modulation $f(t)$,
while $D_{\langle x\rangle}$ can be made even smaller using the feedback.

\subsection{Packet width squeezing}

        Let us use the natural normalization of $D_x$ and $D_p$ by the ground
state parameters, $d_x\equiv D_x/(\hbar/2m\omega_0)$,
$d_p\equiv D_p/(\hbar m\omega_0/2)$, and similarly
$d_{xp}\equiv D_{xp}/(\hbar/2)$. Then Eqs.\ (\ref{eq-mDx})--(\ref{eq-mDxp})
can be rewritten as
        \begin{eqnarray}
&& \hspace{0.5cm}
\dot{d}_{x}/\omega_0 =  2 d_{xp}
- {\cal C}_0\,|f(t)|\, d_x^2 \, ,
\label{eq-mDx-dles}\\
&& \hspace{0.5cm}
\dot{d}_{p}/\omega_0 = \ -2 d_{xp} +
({\cal C}_0/\eta) \,  |f(t)|- {\cal C}_0 \, |f(t)|\, d_{xp}^2 \, ,
\label{eq-mDp-dles}\\
&& \hspace{0.5cm}
\dot{d}_{xp}/\omega_0 =  d_{p} -d_{x}-
{\cal C}_0 \, |f(t)|\, d_x d_{xp} \, .
\label{eq-mDxp-dles}
\end{eqnarray}

        We have analyzed these equations numerically for the harmonic
(\ref{harm-mod}) and stroboscopic (\ref{strob-mod}) modulation $f(t)$
for several values of the maximum coupling ${\cal C}_0$, concentrating on
the range ${\cal C}_0\leq 1$.
Notice that for the stroboscopic modulation the evolution during each
period of modulation can be calculated analytically using Riccati equations
\cite{DohertyJacobs} that significantly simplifies the numerical
calculations.
 As anticipated, we have found that irrespectively
of the initial conditions, Eqs.\ (\ref{eq-mDx-dles})--(\ref{eq-mDxp-dles})
approach the asymptotic solutions which oscillate with the modulation
frequency $\omega$. Even for small coupling, ${\cal C}_0 \ll 1$, the
asymptotic oscillations are significant in the case of resonance:
$\omega \simeq 2\omega_0$ for harmonic modulation and
$\omega \simeq 2\omega_0 /n$ for the stroboscopic modulation (notice that
at ${\cal C}_0=0$ the variances oscillate with frequency $2\omega_0$).
During the oscillation period the asymptotic solution for $d_x (t)$ reaches
the values
both above and below the stationary solution for $f(t)=1$ which
is \cite{DohertyJacobs,Hopkins}
$d_x=(\sqrt{2}/{\cal C}_0) [(1+{\cal C}_0^2/\eta)^{1/2} -1]^{1/2}$
 and becomes $d_x=1/\sqrt{\eta}$
for ${\cal C}_0\ll 1$.
    Most importantly, the squeezed state, $d_x < 1$, may be achieved
for both harmonic and stroboscopic modulation (momentum squeezing is also
achieved; however, we do not analyze it in this paper).

\begin{figure}
\centerline{ \epsfxsize=2.8in \hspace{-0.4cm}
\epsfbox{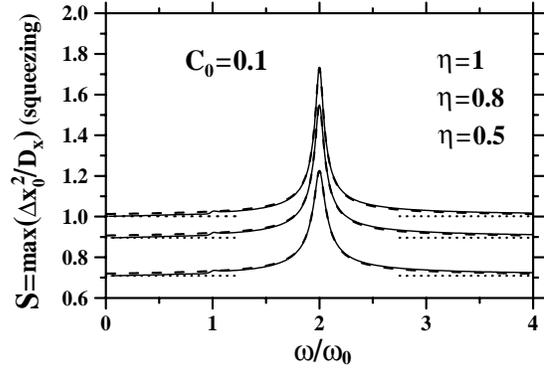}
}
\vspace{0.2cm}
\caption{Dependence of the packet width squeezing ${\cal S}$
(maximized over the modulation period) on the frequency $\omega$
of the harmonic modulation (\protect\ref{harm-mod}) of the measurement
strength, for three values of the quantum efficiency $\eta$ of the detector.
Solid lines show the numerical results, dashed lines are the analytical
results corresponding to Eqs.\ (\protect\ref{S-A}) and
(\protect\ref{Sq-A-cos}), and dotted lines are the asymptotes
${\cal S}=\sqrt{\eta}$.
}
\label{Fig1}
\end{figure}

        Figure 1 shows the maximum squeezing over the oscillation period
for the asymptotic solution, ${\cal S}=\mbox{max}_t [1/d_x(t)] =
\mbox{max}_t [\Delta x_0^2/D_x(t)]$, as a function of the modulation
frequency $\omega$ for the harmonic modulation (\ref{harm-mod})
in the case of weak coupling ${\cal C}_0=0.1$. One can see that
for the ideal detector, $\eta =1$, the squeezing ${\cal S}\approx 1.73$
is achieved at $\omega =2\omega_0$ and decreases to ${\cal S} \approx 1$
(which corresponds to the ground state width) away from the resonance.
The resonances at $\omega =2\omega_0/n$, $n\geq 2$, are barely visible
and lead to small shoulders rather than to peaks.
For nonideal detectors, $\eta <1$, the height of the resonance peak
decreases, ${\cal S}(2\omega_0)\approx 1.73 \sqrt{\eta}$,
while the width increases; the squeezing becomes impossible, ${\cal S} <1$,
at $\eta < 1/3$.

        Much stronger squeezing of the packet width can be achieved
for the stroboscopic modulation (\ref{strob-mod}). (Correspondingly,
the efficiency $\eta$ can also be much lower -- see analytics below.)
 Figure 2 shows
${\cal S}(\omega )$ for the ideal detector with ${\cal C}_0=0.5$
and pulse duration $\delta t=0.05 T_0$,
where $T_0=2\pi /\omega_0$ is the nanoresonator period. One can see that
the sharp resonances at $\omega =2\omega_0/n$ have equal height; however,
their width decreases with $n$. [If modulation (\ref{strob-mod}) is
modified so that $f(t)=\mbox{const}>0$ during ``off'' phase, then the peak
height also decreases with $n$.] For smaller coupling ${\cal C}_0$ the
peak height remains practically the same, but the peak width decreases;
this is the reason why we chose relatively large coupling in Fig.\ 2
in order to have a noticeable peak width. For smaller pulse duration
$\delta t$, the squeezing peaks in Fig.\ 2 would become higher and narrower,
while  the detector nonideality makes peaks lower and wider
(this will be evident from the analytical results discussed below).

\begin{figure}
\centerline{ \epsfxsize=2.9in \hspace{-0.4cm}
\epsfbox{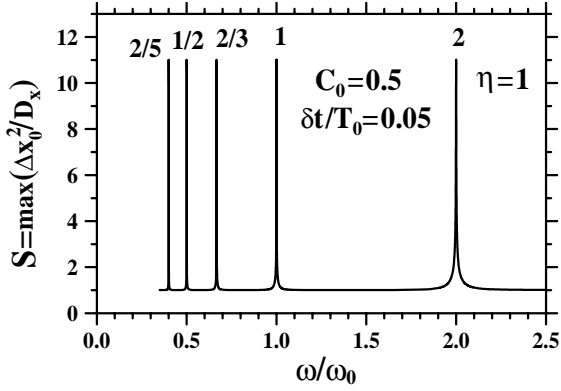}
}
\vspace{0.2cm}
\caption{Numerical results for the packet width squeezing ${\cal S}$
as a function of modulation
frequency $\omega$ for the stroboscopic measurement modulation
(\protect\ref{strob-mod}) with the pulse width $\delta t$. Efficient
squeezing occurs at $\omega \approx 2\omega_0/n$.
The height of the squeezing peaks is proportional to
$\sqrt{\eta}(\delta t)^{-1}$
[Eq.\ (\protect\ref{S-peak-str})] while their width is proportional to
${\cal C}_0(\delta t)^3/n^2\sqrt{\eta}$ [Eq.\ (\protect\ref{S-width-str})].
}
\label{Fig2}
\end{figure}

\subsubsection{Evolution of the state purity}

        Before discussing the analytical results for squeezing, let us
briefly discuss the evolution of the state purity,
$\mbox{Tr}(\rho^2) =(\hbar /2)/\sqrt{D_xD_p-D_{xp}^2}=1/\sqrt{u}$,
where $u=d_x d_p-d_{xp}^2$.
From Eqs.\ (\ref{eq-mDx-dles})--(\ref{eq-mDxp-dles}) it is easy to derive
the equation
        \begin{equation}
    \dot{u} = \omega_0 {\cal C}_0 |f(t)| d_x (\eta^{-1} -u).
        \end{equation}
Since ${\cal C}_0$ and $d_x$ are positive, the asymptotic solution
of this equation is obviously $u=1/\eta$ and therefore the state purity
reaches the asymptote $\mbox{Tr}(\rho^2)= \sqrt{\eta}$. In particular,
in the case of ideal detector, $\eta =1$, the state eventually becomes
pure (similar to the case of a qubit measurement \cite{Kor-99-01}).
It is interesting to note that the typical purification time is
comparable to the time of reaching the asymptotic regime.

\subsubsection{Analytics for harmonic modulation}

        Without measurement, ${\cal C}_0 =0$,
Eqs.\ (\ref{eq-mDx-dles})--(\ref{eq-mDxp-dles}) have the solution
        \begin{eqnarray}
&& d_x(t)=\sqrt{\eta^{-1}+A^2} - A\,\cos(2\omega_0 t + \varphi) \, ,
\label{dx-str}\\
&& d_p(t)=\sqrt{\eta^{-1}+A^2} + A\,\cos(2\omega_0 t + \varphi) \, ,
\label{dp-str}\\
&& d_{xp}(t)=A\,\sin(2\omega_0 t + \varphi) \, ,
        \label{dxp-str}
        \end{eqnarray}
with arbitrary amplitude $A$ and phase $\varphi$. (Notice that these
equations satisfy the condition $u=1/\eta$.) For weak coupling,
${\cal C}_0/\eta \ll 1$, and harmonic modulation (\ref{harm-mod}) in the
vicinity of the resonance, $\omega \simeq 2\omega_0$,
it is natural to look for the asymptotic solution of
Eqs.\  (\ref{eq-mDx-dles})--(\ref{eq-mDxp-dles}) in the form
(\ref{dx-str})--(\ref{dxp-str}) with $2\omega_0$ replaced with $\omega$
(actually, $A$ and $\varphi$ vary in time with frequency $\omega$, but
variations are negligible at ${\cal C}_0/\eta \ll 1$).

        To find $A$ and $\varphi$, we substitute these equations into the
equation
$\int_{-\pi/\omega}^{\pi/\omega} f(t) (\eta^{-1} - d_x^2 - d_{xp}^2) \,
dt = 0$ which follows from the stationarity condition,
$\int_{-\pi/\omega}^{\pi/\omega} (\dot{d}_{x} + \dot{d}_{p})\, dt =0$,
and Eqs.\ (\ref{eq-mDx-dles})--(\ref{eq-mDp-dles}). This gives us
the relation
        \begin{equation}
A = \frac{1}{2}\,\sqrt{\eta^{-1} + A^2}\, \cos{\varphi} .
        \label{A-cosine}
        \end{equation}
We find numerically that $\varphi =0$ at the resonance, $\omega = 2\omega_0$.
(This is quite natural, corresponding to smaller $d_x$ at
larger measurement strength, and is also proven below). Then from
Eq.\ (\ref{A-cosine}) we find $A=1/\sqrt{3\eta}$ and therefore
        \begin{equation}
{\cal S} (2\omega_0) = \sqrt{3\eta}
        \label{S-max-harm}
        \end{equation}
since the maximum squeezing ${\cal S}$ and the amplitude $A$ are related
as
        \begin{equation}
{\cal S}= \eta ( A+\sqrt{A^2+\eta^{-1}} ) .
        \label{S-A}
        \end{equation}
This result confirms the numerical result for the peak height in Fig.\ 1.

        To find the shape of the resonant peak, we need one more
equation relating $A$ and $\varphi$. It can be obtained by deriving equation
for $\ddot{d}_{xp} (t)$ from Eqs.\ (\ref{eq-mDx-dles})--(\ref{eq-mDxp-dles}),
and equating the $\sin (\omega t +\varphi )$ component for the two sides of
the
equation (assuming ${\cal C}_0/\eta \ll 1$ and $\omega \approx 2\omega_0$).
In this way we obtain
        \begin{equation}
(4 \omega_0^2 - \omega^2)\,A  =
 \eta^{-1} {\cal C}_0\,\omega_0^2 \sin{\varphi} \, .
         \label{2eqn-cos}
        \end{equation}
In particular, this proves that $\varphi =0$ at $\omega =2\omega_0$.
Combining Eqs.\ (\ref{A-cosine}) and (\ref{2eqn-cos}) we find the amplitude
$A$ as
        \begin{equation}
A(\omega) = \sqrt{ \frac{2/\eta }{3+g(\omega) +
\sqrt{g^2(\omega) + 10 g(\omega) +9}} } \, ,
        \label{Sq-A-cos}
        \end{equation}
where $ g(\omega) = 16\eta (2-\omega/\omega_0)^2/{\cal C}_0^2$.
This result gives us the analytical expression for squeezing ${\cal S}$ via
Eq.\ (\ref{S-A}). The corresponding squeezing is shown by the dashes lines
in Fig.\ 1, which practically coincide with the solid lines representing
the numerical results.
Notice that the linewidth of the peak is proportional to
${\cal C}_0/\sqrt{\eta}$;
 away from the resonance $A$ decreases to zero, and the squeezing
approaches ${\cal S}=\sqrt{\eta}$, which is the same as for the case
without modulation \cite{DohertyJacobs}. The analytical result for
${\cal S}(\omega )$ works well for coupling ${\cal C}_0$ up to approximately
0.3.
        It is curious that rather complex shape of the resonance peak
given by Eqs.\ (\ref{S-A}) and (\ref{Sq-A-cos}) is quite close to the square
root of the Lorentzian shape:
        \begin{equation}
{\cal S} (\omega ) \approx \sqrt{\eta } \left( 1 +
\frac{\sqrt{3}-1}
{ \sqrt{1 + [ (\omega-2 \omega_0)/\Delta\omega ]^2}} \right)
        \label{sqrt-Lorenz}
        \end{equation}
with $\Delta\omega \simeq 0.36\, \omega_0\, {\cal C}_0/\sqrt{\eta}$.

\subsubsection{Analytics for stroboscopic modulation}

        In the case of stroboscopic modulation (\ref{strob-mod}) of the
measurement strength, the variances $d_x$, $d_p$ and $d_{xp}$ should
follow Eqs.\ (\ref{dx-str})--(\ref{dxp-str}) during the ``off'' phase of
the modulation, while during the measurement pulse of duration
$\delta t$ (``on'' phase) the parameters $A$ and $\varphi$
slowly change (we again assume the weak coupling limit) in accordance with
Eqs.\ (\ref{eq-mDx-dles})--(\ref{eq-mDxp-dles}). In particular,
close to the $n$th resonant peak of Fig.\ 2, $\omega \approx 2\omega_0/n$,
the phase $\varphi$ should change during the pulse by the small amount
        \begin{equation}
\delta \varphi = -2\omega_0\,\frac{2\pi}{\omega} + 2\pi\,n \approx
 \pi\, n^2\, (\omega/\omega_0 - 2/n)
        \label{deltphi}
        \end{equation}
 in order to match $2\pi/\omega$ periodicity of the asymptotic
solution with the periodicity of free oscillations
(\ref{dx-str})--(\ref{dxp-str}). On the other hand, $\delta \varphi$ can
be found from the equation
        \begin{equation}
\dot{\varphi} =-4\omega_0 {\cal C}_0 \eta^{-1} |f(t)| d_{xp}/
        [(d_p-d_x)^2+4d_{xp}^2]
        \label{dot-phi}
        \end{equation}
which follows from from Eqs.\ (\ref{eq-mDx-dles})--(\ref{eq-mDxp-dles}).
Integrating Eq.\ (\ref{dot-phi}) within the pulse interval
$|t|\leq \delta t/2$ using Eqs.\ (\ref{dx-str})--(\ref{dxp-str}) in which
 $A$ and $\varphi$ are assumed constant, we obtain
$\delta \varphi = - {\cal C}_0 \sin (\omega_0\delta t)/\eta A$.
Combining this result with Eq.\ (\ref{deltphi}) we obtain an equation
relating $A$ and $\varphi$:
        \begin{equation}
\pi n^2 A (\omega/\omega_0 - 2/n) = \eta^{-1}{\cal C}_0
\sin (\omega_0 \delta t)\, \sin \varphi \, .
        \label{2eqn}
        \end{equation}

To obtain one more equation for $A$ and $\varphi$, we use the condition
$\int_{-\delta t/2}^{\delta t/2} (\dot{d}_x+\dot{d}_p) \, dt =0$.
Expressing the derivative $\dot{d}_x+\dot{d}_p$ from Eqs.\
(\ref{eq-mDx-dles})--(\ref{eq-mDp-dles}) and using Eqs.\
(\ref{dx-str})--(\ref{dxp-str}) we get the equation
        \begin{equation}
A\, \omega_0\,\delta t = \sqrt{\eta^{-1}+A^2}\ \sin(\omega_0 \delta t)\,
\cos{\varphi} \, .
        \label{1eqn}
        \end{equation}
Equations (\ref{2eqn}) and (\ref{1eqn}) are sufficient to find $A$
for the $n$th resonance, though the expression is quite long:
        \begin{equation}
A^2(\omega) = \frac{2\, \eta^{-1}\, \sin^2 (\omega_0 \delta t)}
{B(\omega) + \sqrt{B^2(\omega) + 4\, \tilde{g}(\omega) \sin^2
(\omega_0 \delta t)} }
\label{res-shape-str} ,
\end{equation}
where $B(\omega)=\tilde{g}(\omega ) + (\omega_0 \delta t)^2 - \sin^2
(\omega_0 \delta t)$
and $\tilde{g}(\omega )=\pi^2 n^2 (2/n-\omega/\omega_0)^2 \eta/{\cal C}_0^2$.
The squeezing ${\cal S}$ is obtained from this result using Eq.\ (\ref{S-A}).
The corresponding analytical curves are plotted in Fig.\ 3 by the
dashed lines which practically coincide with the numerical results shown
by the solid lines. One can see that the analytics works well even
for ${\cal C}_0=1$, even though it was assumed ${\cal C}_0\ll 1$ for
the derivation.

\begin{figure}
\centerline{ \epsfxsize=2.8in \hspace{-0.4cm}
\epsfbox{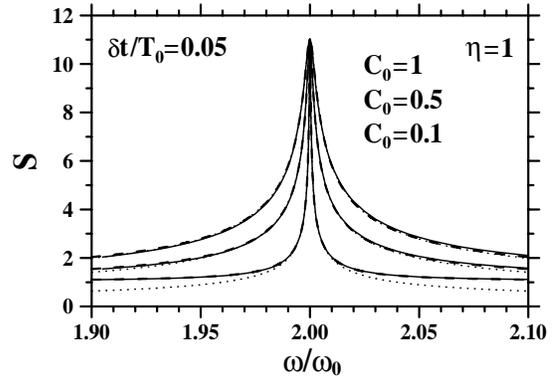}
}
\vspace{0.2cm}
\caption{Squeezing for the stroboscopic modulation for three values
of the coupling with detector ${\cal C}_0$.
Solid lines show numerical results, dashed lines (practically coinciding
with the solid lines) are the analytical results given by Eqs.\
(\protect\ref{res-shape-str}) and (\protect\ref{S-A}), and the dotted
lines are calculated using the simplified equation (\protect\ref{Sq-A-str}).
}
\label{Fig3}
\end{figure}

        The value of squeezing at $\omega =2\omega_0/n$ (peak height) can
be obtained from Eq.\ (\ref{res-shape-str}), but it is easier to use
Eq.\ (\ref{1eqn}) with $\varphi =0$ [which follows from Eq.\ (\ref{2eqn})],
that leads to the result
        \begin{equation}
{\cal S}(2 \omega_0/n) = \sqrt{\eta}\,
\sqrt{\frac{\omega_0 \delta t + \sin (\omega_0 \delta t)}
{\omega_0 \delta t - \sin (\omega_0 \delta t)}} \, .
        \label{Sq-max-strob}
        \end{equation}

        The analytical results significantly simplify in the case of
short pulses, $\delta t \ll T_0=2\pi /\omega_0$, then
        \begin{equation}
A^2(\omega) =
   \frac{6/(\omega_0 \delta t )^2\eta}
{1+\sqrt{1+\left[\frac{6\pi\sqrt{\eta}\, n^2 (\omega-2\omega_0/n) }
{{\cal C}_0 (\delta t)^3 \omega_0^4} \right]^2  }} \, ,
        \label{Sq-A-str}
        \end{equation}
which corresponds to the peak squeezing
        \begin{equation}
{\cal S} (2\omega_0/n) = 2\sqrt{3\eta}/\omega_0\delta t
        \label{S-peak-str}
        \end{equation}
and the full width at half height of ${\cal S} (\omega )$ equal to
        \begin{equation}
\Delta \omega = 4{\cal C}_0 (\delta t)^3\omega_0^4/\pi n^2\sqrt{3\eta}\, .
        \label{S-width-str}
        \end{equation}
The curves calculated using equation (\ref{Sq-A-str}) are shown in Fig.\ 3
by the dotted lines. As one can see, there is a noticeable difference
from the numerical results away from the resonance; however, the main
part of the peak is fitted quite well.

    Equation (\ref{S-peak-str}) shows that the maximum squeezing
in our model does not depend on coupling ${\cal C}_0$. Nanoresonator
interaction with extra environment (for example, due to finite quality
factor) would obviously change this conclusion, because the corresponding
decay of squeezing would compete with the squeezing ``build up'', which rate
is propotional to ${\cal C}_0$ as follows from Eqs.\
(\ref{eq-mDx-dles})--(\ref{eq-mDxp-dles}).

\subsection{Quantum feedback of the packet center}

While the width of the monitored Gaussian packet can be squeezed below the
ground state width as shown in the previous subsection, the center of
the packet undergoes random evolution described by Eqs.\
(\ref{eq-m-x})--(\ref{eq-m-p}), and without feedback diffuses far away
from the origin. For our model which assumes infinite quality factor
of the nanoresonator, the packet center evolves infinitely far away
because the back-action from the detector ``heats up'' the nanoresonator
to formally infinite effective temperature (voltage) of the detector
\cite{DohertyJacobs,MozyrskyMartin,Hopkins}.
  To prevent deviation of the packet center from the origin, we can apply
the quantum feedback described by the force ${\cal F}$ in Eq.\
(\ref{eq-m-p}). Similar to Refs.\ \cite{DohertyJacobs,Hopkins}, we
choose it as
        \begin{equation}
{\cal F} =-m\omega_0 \gamma_x \langle x\rangle - \gamma_p \langle p\rangle .
        \label{fb-force}
        \end{equation}
Notice that the random oscillating evolution of $\langle x\rangle$
and $\langle p\rangle$ can be extracted from the measurement record using
Eqs.\ (\ref{eq-m-x})--(\ref{eq-m-p}) even when the measurement is performed
during only small fraction of the period.

        Following Refs.\ \cite{DohertyJacobs,Hopkins}, we characterize
the distribution of the packet center position $\langle x \rangle $
and center momentum $\langle p\rangle$ by the ensemble averages
(over realizations)
$\langle\langle x \rangle\rangle$ and $\langle\langle p \rangle\rangle$
and the variances $D_{\langle x\rangle}=\langle \langle x\rangle^2\rangle
-\langle\langle x\rangle\rangle^2$,
$D_{\langle p\rangle}=\langle \langle p\rangle^2\rangle
-\langle\langle p\rangle\rangle^2$,
$D_{\langle x\rangle\langle p\rangle}=\langle \langle x\rangle
\langle p\rangle \rangle -\langle\langle x\rangle\rangle
\langle\langle p\rangle\rangle $.
In the notation of doubled angle brackets the inner brackets
mean averaging with individual density matrix $\rho$, while
the outer brackets is averaging over realizations.

        The equations for $\langle\langle \dot{x}\rangle\rangle$
and $\langle\langle \dot{p}\rangle\rangle$ derived from Eqs.\
(\ref{eq-m-x})--(\ref{eq-m-p}) lead to the ensemble averaged position
evolution as
        \begin{equation}
\langle\langle \ddot{x} \rangle\rangle +
\gamma_p \langle\langle \dot{x} \rangle\rangle
+(\omega_0^2+\gamma_x\omega_0)\langle\langle x \rangle\rangle =0,
        \end{equation}
from which it is clear that $\langle\langle x\rangle\rangle$ relaxes to zero
for positive $\gamma_p$.

        Introducing dimensionless variances
$d_{\langle x\rangle}\equiv D_{\langle x \rangle} 2m \omega_0/\hbar$,
$d_{\langle p\rangle}\equiv  D_{\langle p\rangle} 2 /\hbar m \omega_0$, and
$d_{\langle x \rangle\langle p\rangle} \equiv
D_{\langle x\rangle\langle p\rangle} 2/\hbar $,
we derive the following equations from Eqs.\ (\ref{eq-m-x})--(\ref{eq-m-p}):
        \begin{eqnarray}
&& \dot{d}_{\langle x\rangle}/ \omega_0 =
2 d_{\langle x\rangle\langle p\rangle} + {\cal C}_0 |f(t)| \, d_x^2 \, ,
\label{eq-cmDx-dles}\\
&& \dot{d}_{\langle p\rangle}/\omega_0 =
-2 d_{\langle x\rangle\langle p\rangle}
- 2 \mu F d_{\langle x\rangle\langle p\rangle} - 2 F d_{\langle p\rangle}
        \nonumber \\
&&\hspace{1.5cm}  + {\cal C}_0 |f(t)|\, d_{xp}^2 \, ,
\label{eq-cmDp-dles}\\
&& \dot{d}_{\langle x\rangle\langle p\rangle}/\omega_0 =
 d_{\langle p\rangle} - d_{\langle x\rangle}
- \mu F d_{\langle x\rangle} - F d_{\langle x\rangle\langle p\rangle}
\nonumber\\
&&\hspace{1.5cm}
+ {\cal C}_0  |f(t)|\, d_x d_{xp}\, ,
        \label{eq-cmDxp-dles}
        \end{eqnarray}
where $F= \gamma_p/\omega_0$ and $\mu = \gamma_x/\gamma_p$
are the dimensionless feedback parameters.

        We have simulated these equations numerically using the asymptotic
solutions of Eqs.\ (\ref{eq-mDx-dles})--(\ref{eq-mDxp-dles}) for $d_x$,
$d_p$, and $d_{xp}$. We have mostly studied
the resonance $\omega =2\omega_0$ in the weakly coupling regime.
The main finding is that for both harmonic
(\ref{harm-mod}) and stroboscopic (\ref{strob-mod}) modulation of measurement,
 the center
position variance $d_{\langle x\rangle}$ can be made much smaller than
the packet variance $d_x$ at time moments $t=j2\pi/\omega$ when the packet
squeezing is at maximum. Therefore, the corresponding worsening of the
ensemble-averaged squeezing defined as $(d_x+d_{\langle x\rangle}+
\langle\langle x\rangle\rangle^2/\Delta x_0^2)^{-1}$ is negligible.

As an example, for the stroboscopic modulation with $\delta t/T_0=0.05$ and
${\cal C}_0=0.1$ at $\omega =2\omega_0$, the ratio
$d_{\langle x\rangle}/d_x$ at $t=jT_0/2$ is around 0.4\%
for $\mu =0$ and $F=0.5$, and around 0.1\% for $\mu =10$ and $F\gg 1$.
[The term with $\gamma_x$ in Eq.\ (\ref{fb-force}) is not necessary;
however, it improves squeezing of the packet center, so nonzero $\mu$
is beneficial.] The ratio $d_{\langle x\rangle}/d_x$ decreases with
decrease of the pulse width $\delta t$ and decrease of coupling ${\cal C}_0$.
It is important to notice that $d_{\langle x\rangle}$ scales linearly with
${\cal C}_0$ as follows from Eqs.\
(\ref{eq-cmDx-dles})--(\ref{eq-cmDxp-dles}), while $d_x$ as well as
$d_p$ and $d_{xp}$ do not depend on ${\cal C}_0$ at $\omega =2\omega_0$
and ${\cal C}_0/\eta \ll 1$ (this was also checked numerically).
Therefore the ratio $d_{\langle x\rangle}/d_x$ can be made arbitrary
small at small coupling.

        The analytical results (which will be described in more detail
elsewhere) show that in the case $\mu\omega_0\delta t\gg 1$ and $F\gg 1$,
the variance of the packet center at the middle of the measurement pulse
is $d_{\langle x\rangle} ={\cal C}_0 (\omega_0 \delta t)^2/24\mu\eta$,
which can be obviously made much smaller than ${\cal S}^{-1}$ given by
Eq.\ (\ref{S-peak-str}) at sufficiently small ${\cal C}_0/\sqrt{\eta}$,
$\omega_0\delta t$, and/or sufficiently large $\mu$.

\subsection{Observability of the squeezed state}

        The fact that the squeezed state of a nanoresonator can be prepared
by the modulated measurement and quantum feedback, does not automatically
mean that this state may be useful for the measurement of extremely weak
forces, and even that such state can be checked experimentally in a
straightforward way. As an example of such problem, in one of setups analyzed
in Ref.\ \cite{Wiseman-94} the squeezed in-loop optical state is realized
by using quantum feedback, but the squeezing of the output light is
impossible.

We have studied the possibility to verify the squeezed state of the
nanoresonator in the following way. After the preparation of the squeezed
state by stroboscopic measurement and feedback, the feedback
at some moment ($t=0$) is switched off, while the stroboscopic measurement
continues. Considering for simplicity the case of one measurement per
nanoresonator period ($n=2$, $\omega =\omega_0$), we calculate the average
of the position measurements (each pulse gives a very imprecise measurement
because of weak coupling):
        \begin{equation}
X_N = \frac{1}{N} \sum_{j=1}^N \frac{1}{\delta t \, k_0}
 \int_{jT_0-\delta t/2}^{jT_0+\delta t/2} [I(t)-I_{00}] \, dt \, .
        \label{X_N}
        \end{equation}
The idea is that for a squeezed initial state, $X_N$ can be much closer
to zero than if we would start with the ground state.

        The analysis of the distribution of $X_N$ (over realizations)
is very simple in the case of instantaneous but imprecise measurements,
$\delta t\rightarrow 0$, ${\cal C}_0\delta t = \mbox{const}$, since the
Hamiltonian evolution of the resonator in between the measurements can be
completely neglected, and therefore $N$ measurements are equivalent to
one $N$-times stronger measurement. This gives us the variance of $X_N$
equal to
        \begin{equation}
D_{X,N} = \frac{\hbar}{2m\omega_0} \left( \frac{1}{\cal S} +
        \frac{1}{N{\cal C}_0 \omega_0 \delta t} \right)  ,
        \label{DXN}
        \end{equation}
where the first term is due to the initial packet width, while the second
term is the inaccuracy of the measurement which improves with $N$.
Obviously, at $N \gg 1/{\cal C}_0\omega_0\delta t$ this variance for a
squeezed
state (${\cal S}>1$) is significantly different from the variance for the
ground state (${\cal S}=1$). Even though this difference can be
rigorously verified only by performing many experiments to accumulate
statistics
for $D_{X,N}$, it can be observed even in the single experiment with
good reliability if ${\cal S}\gg 1$. (In a single realization
the failure probability for distinguishing squeezed and ground states is
crudely ${\cal S}^{-1/2}$.)

        Unfortunately, this result requires the assumption of infinitely
strong coupling, so it is not obvious if it holds in the case of weak
coupling or not. The possible problem is that for sufficiently large $N$
which makes the second term in Eq.\ (\ref{DXN}) sufficiently small,
the nanoresonator
heating due to measurement back-action may already eliminate the squeezing
(the feedback is off). We have calculated $D_{X,N}$ for stroboscopic
modulation numerically using Eqs.\ (\ref{eq-m-x})--(\ref{eq-m-p})
and have found that there is still a range of $N$
where the squeezed and ground initial states lead to significantly different
$D_{X,N}$ and therefore can be reliably distinguished.
As an example, for ${\cal C}_0=0.1$, $\eta=1,$ and $\delta t=0.02 T_0$,
the normalized
variance $d_{X,N}\equiv D_{X,N}2m\omega_0/\hbar$ achieves a minimum of 0.078
(at $N\simeq 4\times 10^3$),
which is crudely two times larger than the contribution from the initial
squeezing $1/{\cal S}=0.036$, and is still significantly smaller than
the ground state limit $d_{X,N}\geq 1$. We have found numerically that
the minimum $d_{X,N}$ scales linearly with the pulse width $\delta t$
(similarly to $1/{\cal S}$)
and practically does not depend on coupling ${\cal C}_0$ at
${\cal C}_0\leq 1$. This hints that the product
${\cal S}\times \mbox{min}_N d_{X,N}$ is practically a constant approximately
equal to 2, and therefore verification of the squeezed state by a weakly
coupled detector is almost as efficient as the verification by instantaneous
measurements (within a factor of about 2).

\section{Conclusion}

        In this paper we have shown that the uncertainty of the
nanoresonator position can be squeezed significantly below the ground state
level by the modulated in time measurement of the nanoresonator position
with the QPC or RF-SET detector. The measurement strength is modulated
by applying the periodic voltage across the detector. The mechanism of
squeezing is similar to the QND measurements \cite{BraginskyKhalili},
though a significant difference in our case is the continuous measurement
with weak coupling to the detector.
We have considered harmonic (\ref{harm-mod}) and stroboscopic
(\ref{strob-mod}) modulations and found that only a moderate squeezing
${\cal S}\leq \sqrt{3\eta}$ is possible for the harmonic modulation with
frequency $\omega \approx 2\omega_0$
[see Eqs.\ (\ref{S-max-harm}) and (\ref{Sq-A-cos}), and Fig.\ 1]. However,
the stroboscopic modulation can lead to an arbitrary strong squeezing
${\cal S} \leq 2\sqrt{3\eta}/\omega_0\delta t$
for sufficiently short measurement pulses $\delta t$ applied with frequency
$\omega =2\omega_0/n$ [see Eqs.\
(\ref{Sq-max-strob})--(\ref{S-width-str}) and Fig.\ 2]. Obviously,
the state width oscillates with time, so that the maximum position squeezing
is achieved periodically (with period close to $\pi/\omega_0$), while
the maximum squeezing of momentum happens with
$\pi/2\omega_0=T_0/4$ time shift (when ${\cal S}<1$).

        While the modulated measurement squeezes the width of the
state (packet), the position of the packet center $\langle x\rangle$
fluctuates due to
random back-action from the detector; so to keep the packet center near
$x=0$ we need to apply quantum feedback. We have found that the feedback
can keep the deviation of $\langle x\rangle$ from zero much smaller
than the packet width, which means that the ensemble-averaged squeezing
practically does not differ from the packet width squeezing.

        In this paper we have used the Bayesian formalism \cite{Kor-99-01}
for the description of the quantum measurement process. However,
since the obtained equations practically coincide with the equations
used in Refs.\ \cite{DohertyJacobs,Hopkins}, we have followed those
papers to a large extent, especially for the analysis of the evolution
of the Gaussian states.

        An important issue is the possibility to use the squeezed states
of the nanoresonator for the measurement of weak forces with the accuracy
beyond the standard quantum limit. Even though we did not consider this
question explicitly, we have found that the state squeezing can be verified
(with high
reliability) even in a single measurement run by a weakly coupled detector.

        The main drawback of the present theory is the assumption of
a very large quality factor $Q$ of the nanoresonator. Crude preliminary
analysis indicates that our results for the stroboscopic modulation
 are valid for $Q \gg {\cal S}^2/{\cal C}_0\sqrt{\eta} \sim
\sqrt{\eta}/{\cal C}_0(\omega_0\delta t)^2$ and sufficiently
small temperature of the environment, $T/\hbar\omega_0 \ll
\sqrt{\eta} Q{\cal C}_0/{\cal S}^3 \sim  Q{\cal C}_0
(\omega_0\delta t)^3/\eta$.  These conditions seem to be within
present-day experimental reach for moderate squeezing.

\section*{Acknowledgment}
The authors would like to thank D. Averin, A. Doherty, S. Habib,
K. Jacobs, K. Likharev, I. Martin, and G. Milburn
for fruitful discussions and remarks.
This work was supported by NSA and ARDA under the ARO grant
DAAD 19-01-1-0491 (R.R.\ and A.N.K.) and by the National Security
Agency (K.S.).


\begin{thebibliography}{99}


\bibitem{BraginskyKhalili}
        V.~B. Braginsky and F.~Ya. Khalili,
        {\it Quantum measurement},  (Cambridge Univ. Press, Cambridge, 1992).

\bibitem{Braginsky2}
        V.~B. Braginsky, Yu.~I. Vorontsov, and F.~Ya. Khalili,
        JETP\ Lett. {\bf 27}, 276 (1978).

\bibitem{Thorne}
        K.~S. Thorne, R.~W.~P. Drever, C.~M. Caves, M. Zimmermann, and
        V.~D. Sandberg, Phys.\ Rev.\ Lett. {\bf 40}, 667 (1978).

\bibitem{CavesRevModPhys}
        C.~M. Caves, K.~S. Thorne, R.~W.~P. Drever, V.~D. Sandberg, and
        M. Zimmermann, Rev. Mod. Phys. {\bf 52}, 341 (1980).

\bibitem{Kimble} H. J. Kimble, Y. Levin, A. B. Matsko, K. S. Thorne, and
        S. P. Vyatchanin, Phys.\ Rev. D {\bf 65}, 022002 (2002).

\bibitem{Braginsky-2003} V. B. Braginskii, Physics-Uspekhi {\bf 46}, 81 (2003).

\bibitem{Averin} D. V. Averin, Phys.\ Rev.\ Lett. {\bf 88}, 207901 (2002).

\bibitem{Bulaevskii} L. Bulaevskii, M. Hruska, A. Shnirman, D. Smith, and
        Y. Makhlin, Phys.\ Rev.\ Lett. {\bf 92}, 177001 (2004).

\bibitem{Geremia} JM Geremia, J. K. Stockton, and H. Mabuchi, Science
        {\bf 304}, 270 (2004).

\bibitem{ClelandRoukes}
        A.~N. Cleland and M.~L. Roukes,
         Appl.\ Phys.\ Lett. {\bf 69}, 2653 (1996);
    A.~N. Cleland and M.~L. Roukes, Nature {\bf 392}, 160 (1998);
    E. Buks and M. L. Roukes, Europhys.\ Lett. {\bf 54}, 220 (2001).

\bibitem{Craighead} D. W. Carr and H. G. Craighead,
        J.\ Vacuum Sci.\ Tech. B {\bf 15}, 2760 (1997);
     H. G. Craighead, Science {\bf 290}, 1532 (2000);
    M. Zalalutdinov, B. Ilic, D. Czaplewski, A. Zehnder, H.~G. Craighead,
        and J.~M. Parpia, Appl.\ Phys.\ Lett. {\bf 77}, 3287, (2000).

\bibitem{Roukes-1GHz} X. Ming, H. Huang, C.~A. Zorman, M. Mehregany, and
        M.~L. Roukes, Nature {\bf 421}, 496 (2003).

\bibitem{Cleland-2003} R. G. Knobel and A. N. Cleland, Nature {\bf 424}, 291 (2003).

\bibitem{Schwab-QL} M.~D. LaHaye, O. Buu, B. Camarota, and K.~C. Schwab,
        Science {\bf 304}, 74 (2004).

\bibitem{Cleland-QPC} A.~N. Cleland, J.~S. Aldridge, D.~C. Driscoll, and
        A.~C. Gossard, Appl.\ Phys.\ Lett. {\bf 81}, 1699 (2002).

\bibitem{Blencowe} M. Blencowe, Phys.\ Reports {\bf 395}, 159 (2004);
   M. P. Blencowe and M. N. Wybourne, Appl.\ Phys.\ Lett. {\bf 77}, 3845 (2000);
   A. D. Armour, M. P. Blencowe, and Y. Zhang, Phys.\ Rev. B {\bf 69}, 125313 (2004).

\bibitem{MozyrskyMartin} D. Mozyrsky and I. Martin,
         Phys.\ Rev.\ Lett. {\bf 89}, 018301 (2002).


\bibitem{ArmourBlencoweSchwab}
        A.~D. Armour, M.~P. Blencowe, and K.~C. Schwab,
         Phys. Rev. Lett. {\bf 88}, 148301 (2002).


\bibitem{Smirnov} A. Y. Smirnov, L. G. Mourokh, and N. J. M. Horing,
         Phys. Rev. B {\bf 67}, 115312 (2003).

\bibitem{Schwab} K. Schwab,
         Appl.\ Phys.\ Lett. {\bf 80}, 1276 (2002).


\bibitem{Knobel} R. Knobel and A.~N. Cleland,
         Appl.\ Phys.\ Lett. {\bf 81}, 2258 (2002).


\bibitem{Hopkins} A. Hopkins, K. Jacobs, S. Habib, and K. Schwab,
         Phys.\ Rev. B {\bf 68}, 235328 (2003).


\bibitem{ShnirmanMartin} I. Martin, A. Shnirman, L. Tian, P. Zoller,
         Phys.\ Rev. B {\bf 69}, 125339 (2004).

\bibitem{Ruskov-fb} R. Ruskov and A. N. Korotkov,
          Phys.\ Rev. B {\bf 66}, 041401 (2002);
          A. N. Korotkov, cond-mat/0404696 (2004).

\bibitem{Wiseman} H. M. Wiseman and G. J. Milburn,
                Phys.\ Rev.\ Lett. {\bf 70}, 548 (1993).

\bibitem{Blencowe-sq} M. P. Blencowe and M. N. Wybourne,
                    Physica B {\bf 280}, 555 (2000).

\bibitem{RugarGrutter} D. Rugar and P. Gr\"{u}tter,
         Phys.\ Rev.\ Lett. {\bf 67}, 699 (1991).


\bibitem{Gurvitz} S.~A. Gurvitz,
         Phys.\ Rev. B {\bf 56}, 15215 (1997).


\bibitem{Kor-Av} A.~N. Korotkov and D. V. Averin,
                Phys.\ Rev. B {\bf 61}, 165310 (2001).

\bibitem{GoanMilburn} H.~S. Goan and G.~J. Milburn,
         Phys. Rev. B {\bf 64}, 235307 (2001).


\bibitem{Kor-nonid} A.~N. Korotkov,
        Phys.\ Rev. B {\bf 67}, 235408 (2003).

\bibitem{Kor-99-01} A.~N. Korotkov,
         Phys.\ Rev. B {\bf 60}, 5737 (1999);
        A.~N. Korotkov, Phys.\ Rev. B {\bf 63}, 115403 (2001).

\bibitem{DohertyJacobs} A.~C. Doherty and K. Jacobs,
         Phys.\ Rev. B {\bf 60}, 2700 (1999).


\bibitem{Mensky} M.~B. Mensky,
        Physics-Uspekhi {\bf 168}, 1017, (1998).

\bibitem{GWM} M. J. Gagen, H. M. Wiseman, and G. J. Milburn,
        Phys.\ Rev. A {\bf 48}, 132, (1993).

\bibitem{Lindblad} G. Lindblad, Comm.\ Math.\ Phys. {\bf 48}, 119 (1976).

\bibitem{Leggett} A.~O. Caldeira and A.~J. Leggett,
        Ann.\ Phys. (N.Y.), {\bf 149}, 374 (1983).

\bibitem{CavesMilburn} C. Caves and G.~J. Milburn,
        Phys.\ Rev. A {\bf 36}, 5543 (1987).

\bibitem{ZurekHabibPaz-prl} W.~H. Zurek, S. Habib, and J.~P. Paz,
        Phys.\ Rev.\ Lett. {\bf 70}, 1187 (1993).


\bibitem{HalliwellZoupas} J. Halliwell and A. Zoupas,
         Phys.\ Rev. B {\bf 52}, 7294 (1995).


\bibitem{BreslinMilburn} J.~K. Breslin and G.~J. Milburn,
         Phys.\ Rev. A {\bf 55}, 1430 (1997).


\bibitem{Gardiner} C.~W. Gardiner,
        {\it Quantum noise},  (Springer, Berlin, 1991).

\bibitem{Man'ko}
        V.~V. Dodonov, E.~V. Kurmyshev, and V.~I. Man'ko,
         Phys.\ Lett. A {\bf 79}, 150 (1980).

\bibitem{Wiseman-94} H. M. Wiseman and G. J. Milburn,
                   Phys.\ Rev. A {\bf 49}, 1350 (1994).
\end{thebibliography}
\end{document}